\documentstyle[11pt]{article}

\newcommand{\sect}[1]
{\setcounter{equation}{0}\setcounter{footnote}{0}\section{#1}}

\newcommand{\foot}[1]{\footnote{\small #1}}

\newcommand{\bea}{\begin{eqnarray}}
\newcommand{\ena}{\end{eqnarray}}

\newcommand{\CMP}[1]{Comm. Math. Phys.{\bf #1}}
\newcommand{\PRD}[1]{Phys. Rev. {\bf D#1}}
\newcommand{\PRL}[1]{Phys. Rev. Lett. {\bf #1}}
\newcommand{\JHEP}[1]{J. High Energy Phys. {\bf #1}}
\newcommand{\PLB}[1]{Phys. Lett. {\bf B#1}}
\newcommand{\ATMP}[1]{Adv. Theor. Math. Phys. {\bf #1}}
\newcommand{\NPB}[1]{Nucl. Phys. {\bf B#1}}
\newcommand{\MPLA}[1]{Mod. Phys. Lett. {\bf A#1}}
\newcommand{\CQG}[1]{Class. Quant. Grav. {\bf #1}}
\newcommand{\AP}[1]{Ann. Phys. {\bf #1}}

\begin{document}
\large

\begin{titlepage}
\begin{flushright}
hep-th/9903065\\
OU-HET 316\\
\end{flushright}
\begin{center}
\vspace{2cm}
{\Large \bf AdS$_{\bf 3}$ Gravitational Instantons from 
Conformal Field Theory}\\
\vspace{2cm}
Shuhei Mano\foot{e-mail: mano@het.phys.sci.osaka-u.ac.jp}\\
\vspace{2cm}
{\it Department of Physics,\\Osaka University\\Osaka 560-0043, Japan}\\
\vspace{4cm}
{\bf Abstract}
\end{center}

A conformal field theory on the boundary of three-dimensional 
asymptotic anti-de Sitter spaces which appear as near horizon geometry 
of D-brane bound states is discussed.
It is shown that partition functions of gravitational instantons
appear as high and low temperature limits of the partition function
of the conformal field theory.
The result reproduces phase transition between the anti-de Sitter space
and the BTZ black hole in the bulk gravity.
\end{titlepage}

\sect{Introduction}\label{intro}

To understand thermodynamics of black holes~\cite{bek,haw}
is important because it provides us some insight of quantum gravity.
The most simple derivation of thermodynamics of black holes is
given by the Euclidian quantum gravity approach~\cite{gibhaw}.
We consider a semiclassical partition function in path-integral
approach to quantize gravity as
\bea
Z_{\rm semi.} = e^{-I_E[g_0]}.\label{semic}
\ena
where $g_0$ is a metric of an Euclidian black hole and $I_E[g_0]$ is 
the Euclidian action.
If we regard this partition function as a statistical mechanical partition
function, we can obtain thermodynamical variables.
The entropy is given as
\bea
S = \frac{A}{4G},\label{behae}
\ena
where $A$ is the area of the event horizon and $G$ is the Newton constant. 
It is natural to think that the entropy of black holes have some 
microscopic origin.
However, the Euclidian quantum gravity approach has no explanation 
about the microscopic origin because of its thermodynamic nature~\cite{mar}.

The microscopic derivation of the entropy of black holes can be given 
in terms of string theory~\cite{strvaf,calmal,malD}.
In this article, we study the D-brane bound states 
which constructed from a D5-brane, a D1-brane on $T^4 \times S^1$.
The D5-brane wraps four-cycle in $T^4$ and $S^1$ $Q_5$ times and
the D1-brane wraps zero-cycle in $T^4$ and $S^1$ $Q_1$ times.
The Kaluza-Klein momentum is added in the direction along $S^1$.
The bound states give five-dimensional Reissner-Nordstr\"om black
holes when compactified on $T^5$.
The thermodynamical entropy is given by 
(\ref{behae}).\foot{This argument holds only for
nonextremal black hole in the Euclidian quantum gravity approach. 
This point will be discussed in section \ref{eucli}.}
On the other hand, the moduli of the bound states can be described 
by the D1-brane worldvolume theory, and it becomes a two-dimensional 
superconformal field theory in low energy.
The superconformal field theory is a nonlinear sigma model whose
target space is a deformation of a symmetric product of 
$k \equiv Q_1Q_5$ copies of $T^4$
\bea
\frac{(T^4)^{\otimes k}}{S(k)},\label{targe}
\ena
where $S(k)$ is a permutation group of $k$ elements.
Because this target space has dimension $4k$,
the superconformal field theory has central charge $c = 6k$.
The level, say, $n$ of the states of the superconformal field
theory is fixed by the Kaluza-Klein momentum.
Because gravity knows only the level, the state has large 
degeneracy in microscopic viewpoint when $n \gg c$.
The entropy becomes the same as (\ref{behae})
for extremal or near extremal black holes.

Recently, it was conjectured that the large $N$ gauge theory on
the boundary of a product space of anti-de Sitter space (AdS) and some 
compact manifold corresponds to supergravity or string 
theory on the bulk~\cite{malN}.
One of the most tractable example of this conjecture is given by
D-brane bound states which was discussed above.
The near horizon geometry of the D-brane bound states become
a product space of the BTZ black hole and $S^3 \times T^4$~\cite{malstrA}.
The BTZ black holes are quotient spaces of $AdS_3$ and reach $AdS_3$ 
asymptotically~\cite{banteizenB}.
It is known that spaces which reach $AdS_3$ asymptotically 
have conformal symmetry at infinity~\cite{brohen},
and its extension to superconformal symmetry is also discussed~\cite{couhen}.
The superconformal field theory on the boundary has central 
charge $c = 3\ell/2G_3 = 6k$, which is the same as that of 
the D1-brane worldvolume theory, where $\ell$ is the
characteristic radius of $AdS_3$ and $G_3$ is the three-dimensional
Newton constant.
Because of this correspondence, we may consider the two conformal field
theories are the same~\cite{malstrA}.
The conjecture is translated into the large $c$ superconformal 
field theory can be described by the bulk gravity and vice versa.
Indeed, the microscopic derivation of entropy of BTZ black holes
is given in terms of this conjecture~\cite{strN}.

Let us recall that the semiclassical partition function (\ref{semic})
has only the contribution of a black hole.
However, in semiclassical approximation,
contributions must come from all metrics of Euclidian solutions
which are consistent with a given boundary condition~\cite{gr}
\bea
Z_{\rm semi.} = \sum_{g_0 \in {\rm inst.}} e^{-I_E[g_0]}.
\ena
These Euclidian solutions are called gravitational instantons.
For example, let us consider a system at an inverse temperature 
$\beta$ contained in asymptotic $AdS_4$~\cite{hawpag}. 
Continued to the Euclidian metric, 
the boundary has topology $S^1 \times S^2$, 
where $S^1$ corresponds to the imaginary time with 
identification.
One of the gravitational instantons which fits this boundary
condition is Euclidian $AdS_4$ whose imaginary time is 
identified with the period $\beta$.
It has topology $S^1 \times {\bf R}^3$, where $S^1$ corresponds to the 
imaginary time.
The Euclidian anti-de Sitter-Schwarzschild black hole at inverse
temperature $\beta$ is another solution which fits the boundary 
condition. It has topology ${\bf R}^2 \times S^2$, 
where ${\bf R}^2$ corresponds to the plane whose angular coordinate
is the imaginary time.

The semiclassical partition function becomes
\bea
Z_{\rm semi.}[\beta]
= Z_{\rm semi.}^{\rm AdS}[\beta] + Z_{\rm semi.}^{\rm BH}[\beta].
\ena
The anti-de Sitter-Schwarzschild black hole has a maximum inverse 
temperature $\beta_0 = 2\pi \ell_4$, where $\ell_4$ is the characteristic 
radius of $AdS_4$.
The anti-de Sitter-Schwarzschild black holes whose inverse temperature is 
lower than $\beta_0$ have positive specific heat and it is stable at
least locally, contrasting to asymptotic flat Schwarzschild black
holes.
The anti-de Sitter-Schwarzschild black holes whose inverse temperature is
lower than $\beta_1 = \sqrt{3}\pi \ell_4$ have lower Euclidian action 
than that of $AdS_4$.
When the inverse temperature is lower than $\beta_1$,
the global minimum of the Euclidian action of the
system comes from the anti-de Sitter-Schwarzschild black
hole\foot{When the inverse temperature becomes so low, the thermal
equilibrium with radiation becomes impossible.}.
We can regard this phenomenon as thermal phase transition between
the anti-de Sitter space and the anti-de Sitter-Schwarzschild black hole.
This phase transition of the bulk gravity can be interpreted as
the large $N$ phase transition, or confinement of the large $N$ gauge 
theory on the boundary~\cite{witH}.

In this article, we discuss the phase transition of
asymptotic $AdS_3$ and construct the partition function of the
conformal field theory on the boundary, which was pointed out 
in~\cite{malstrA}.
The phase transition of the bulk gravity is reproduced in terms of the
partition function of the conformal field theory.

\sect{Euclidian Quantum Gravity}\label{eucli}
\noindent
It is known that the extremal Reissner-Nordstr\"om black hole
has no entropy because of a topological
reason in Euclidian quantum gravity
approach~\cite{hawhorros,teiE,hawhor,gibkal}.
We briefly review the reason and discuss this point for 
the extremal BTZ black hole.

Let us consider thermodynamics of Reissner-Nordstr\"om black
holes. 
The Reissner-Nordstr\"om metric is
\bea
ds^2 = -\left(1-\frac{2GM}{r}+\frac{GQ^2}{r^2}\right) dt^2 +
\left(1-\frac{2GM}{r}+\frac{GQ^2}{r^2}\right)^{-1} dr^2 + 
r^2 d\Omega_2^2,\label{metR}
\ena
where $Q$ is the electric charge of electromagnetic fields and $M$ 
is the ADM mass of the black hole.
The radii of horizons are $r_{\pm}=GM\pm(G^2M^2-GQ^2)^{1/2}$, 
where $r_{\pm}$ are those of the outer and inner horizon 
respectively.
The first law of thermodynamics of Reissner-Nordstr\"om black holes is
\bea
dM = \frac{\kappa}{2\pi}\frac{dA}{4G} + {\varphi}_+ dQ,
\label{1stRN}
\ena
where $\kappa$ is the surface gravity, $\varphi_+$ is the
scalar potential at the event horizon and $A$ is the area of the 
horizon respectively.
We see that $\varphi_+$ may be regarded as the chemical potential.

We write the action in the Hamiltonian form in terms of
ADM formalism~\cite{hawhor,regtei}.
The Euclidian metric is obtained by an analytic continuation of the
metric (\ref{metR}) which is given by the replacement
\bea
t \mapsto -i\tau.
\ena
We identify the imaginary time with a period $\beta$.
We take two boundaries: the one is $S^1 \times S^2$,
where $S^1$ is the imaginary time and $S^2$ is the two-sphere at infinity.
The other is $S^1 \times S^2$, 
where $S^1$ is the imaginary time and $S^2$ is the event horizon.
The Euclidian action for the Euclidian metric over a region $M$, 
which is the bulk of two boundaries, has the form
\bea
I_E = \frac{1}{16\pi G} \int_M d^4x \left(\pi^{ij}\dot{h}_{ij} - 
N^{\bot}{\cal H}_{\bot} - N^i{\cal H}_i\right) + I_{E{\rm matter}} + B,
\ena
where $B$ is the boundary term which is discussed below, and
$I_{E{\rm matter}}$ is the Euclidian action for matter fields.

To discuss thermodynamics of black holes,
we are interested in the variation with respect to 
$M$ and $Q$, because the Euclidian action should be constructed so 
that it has an extremum on the solution.
In the variation, we must fix the thermodynamical conjugate variables.
For the variation being well defined,
we must add the boundary term $B$ to cancel the unwanted boundary terms.
The total boundary terms should be fixed as
\bea
B = \beta M - \beta\varphi_{+} Q - \frac{A}{4G},
\label{bou}
\ena
for the nonextremal black hole.
The first term is needed to the variation at infinity with respect to
$M$ and the second term is needed to the variation at the horizon
with respect to $Q$.
The third term needs more explanation.
The near horizon geometry of the nonextremal black hole 
has topology ${\bf R}^2 \times S^2$,
where $\tau$ is the angular and $(r-r_+)$ is the radial coordinate of 
${\bf R}^2$.
In general a conical singularity sit on the origin of ${\bf R}^2$.
The variation of the Euclidian action at the 
horizon leads~\cite{teiE}
\bea
\frac{1}{4G}\frac{\Theta}{2\pi}\delta A,
\ena
where $\Theta$ is the proper angle at the horizon.
We must fix the boundary term such that the Euclidian action has an extremum on
the solution with no conical singularity at the horizon.
Thus we must add third term in (\ref{bou}).
On the other hand, the near horizon geometry of the 
extremal black hole has topology 
${\bf R} \times S^1 \times S^2$, where $\tau$ is a coordinate of
$S^1$ and there is no conical 
singularity to avoid. This fact means that there is no need to add 
the third term in (\ref{bou}).

The bulk parts of the Euclidian action vanish when constraints hold,
and the Euclidian action has the boundary terms only.
When we identify the semiclassical partition with 
the grand partition function of statistical mechanics,
we obtain the free energy
\bea
F = M - \varphi_+ Q - \frac{1}{\beta}\frac{A}{4G},
\ena
for the nonextremal black hole.
Thus, the entropy of the nonextremal black hole is given as 
(\ref{behae}), and the inverse temperature is 
given as $\beta = 2\pi/\kappa$.
These results are consistent with (\ref{1stRN}).
For the extremal black hole, we obtain free energy
\bea
F = M - \varphi_+ Q,
\ena
and the entropy of the extremal black hole
vanish nevertheless the area exists.
Moreover, the fact that we can identify the imaginary time of the
extremal black hole
in any period because of the topological reason
implies that the extremal black hole can be in
equilibrium with a thermal bath at any temperature.
The fact that the entropy changes discontinuously in 
the extremal limit implies that one should regard a 
nonextremal and an extremal black hole as qualitatively 
different objects~\cite{hawhorros}.
However, as mentioned in section \ref{intro}, the entropy of 
the extremal black hole obeys the same relation (\ref{behae}) 
microscopically as nonextremal black hole.
This discrepancy was discussed in~\cite{sen,hor,ghomit}.

Let us consider thermodynamics of BTZ black holes, which emerges as
the near horizon geometry of the D1-D5-brane bound states~\cite{malstrA}.
The BTZ black hole metric is
\bea
ds^2&=&-N^2dt^2 + N^{-2}dr^2 + r^2\left(d\phi+N^{\phi}dt\right)^2, 
\label{metBTZ}
\ena
where the squared lapse $N^2$ and the angular shift $N^{\phi}$ are
\bea
N^2&=&-8G_3 M+\frac{r^2}{\ell ^2}+\frac{(8G_3 J)^2}{4r^2},\nonumber\\
N^\phi&=&-\frac{8G_3 J}{2r^2},
\ena
where $M$ and $J$ are the mass and angular momentum respectively.
The three-dimensional Newton constant $G_3$ is given by the
compactification and it becomes $\ell/4k$.
The first law of thermodynamics of BTZ black holes is
\bea
dM = \frac{\kappa}{2\pi}\frac{dA}{4G_3} + \Omega_{+}dJ,
\label{1stBTZ}
\ena
where $A$ is the circumference of the event horizon.
$\kappa$ and $\Omega_+$ are the surface gravity and the angular
velocity of the horizon respectively.

The Euclidian metric is obtained by an analytic 
continuation of the metric (\ref{metBTZ}) to imaginary $t$ 
and $J$~\cite{cartei}.
The continuation is given by the replacement
\bea
t&\mapsto&-i\tau,\nonumber\\
J&\mapsto&iJ_E,\label{ancon}
\ena
where $J_E$ is fixed by the Lorentzian angular momentum as
$J_E = J$.
Note that the continuation of the angular momentum to the imaginary 
value is necessary for the metric to be real. 
This is physically sensible, since the angular velocity is now 
a rate of change of the real angle with respect to the imaginary time
\cite{qua}.
The Euclidian metric becomes
\bea
ds^2&=&N^2_E d\tau^2 + N_E^{-2} dr^2 + 
r^2 \left(d\phi+N^{\phi}_Ed\tau\right)^2 \nonumber\\
&=&\frac{(r^2-r^2_+)(r^2-r^2_-)}{r^2 \ell^2} d\tau^2+
\frac{r^2 \ell^2}{(r^2-r^2_+)(r^2-r^2_-)} dr^2
+ r^2\left(d\phi+r_+\frac{ir_-}{r^2} d\tau\right)^2,\nonumber\\
\label{metEBTZ}
\ena
where the squared lapse $N^2_E$ and the angular shift $N^{\phi}_E$ are
\bea
N^2_E&=&-8G_3 M + \frac{r^2}{\ell^2} - \frac{(8G_3 J_E)^2}{4r^2},
\nonumber\\
N^\phi_E&=&- \frac{8G_3 J_E}{2r^2}.
\ena
The radii of horizons become
\bea
r_+ = \left\{\frac{8G_3 M\ell^2}{2}
\left[1 + \sqrt{1 + \frac{J_E^2}{M^2\ell^2}}\right]\right\}^{1/2},
\nonumber\\
r_- = -i|r_-|=\left\{\frac{8G_3 M\ell^2}{2}
\left[1 - \sqrt{1 + \frac{J_E^2}{M^2\ell^2}}\right]\right\}^{1/2}.
\ena
The radius of the outer horizon of the Euclidian black hole is $r_+$. 
It is important that the radius of the inner 
horizon of the Lorentzian black hole is continued to the imaginary value.
This fact means that the Euclidian black hole has only one horizon.
Thus the Euclidian ``extremal'' black hole which is obtained 
from the Lorentzian extremal black hole with the analytic continuation 
(\ref{ancon})
does not have degenerate horizon, namely it has the same topology
${\bf R}^2 \times S^1$ as that of nonextremal Euclidian black holes.

Following the same procedure described for Reissner-Nordstr\"om
black holes, we get the total boundary terms of the Euclidian action as
\bea
B = \beta_E M + \beta_E\Omega_{E+}J_E - \frac{A_E}{4G_3},
\ena
where $\beta_E$ is the period of the imaginary time and 
$\Omega_E$ and $A_E$ are the angular velocity 
and the circumference of the event horizon respectively.
These values are different from the Lorentzian values. 
Moreover, this expression holds for the extremal black hole,
contrasting to Reissner-Nordstr\"om black holes.
The semiclassical partition function becomes
\bea
Z^{\rm BTZ}_{\rm semi} = \exp\left(-\beta_E M - \beta_E\Omega_{E+}J_E + 
\frac{A_E}{4G_3}\right),
\label{parEBTZ} 
\ena
but this expression itself does not reproduce the relation (\ref{1stBTZ}).
Thus, to derive the grand partition function of thermodynamics,
we must apply analytic continuation of the angular momentum $J_E$ 
again to the semiclassical partition function (\ref{parBTZ}) as
\bea
J_E \mapsto -iJ,\label{conba}
\ena
where $J$ is the original Lorentzian angular momentum.
The grand partition function is given as
\bea
Z^{\rm BTZ}
&=&\exp\left(-\beta M + \beta\Omega_+ J + \frac{A}{4G_3}\right)\nonumber\\
&=&\exp\left(\frac{\pi^2\ell^2}{2G_3}\frac{T}{1-\Omega_+^2\ell^2}\right),
\label{parBTZ}
\ena
where $\beta$ is the inverse temperature which is given from $\beta_E$ 
by the continuation (\ref{conba}).
The temperature is given as $T = \kappa/2\pi$ and
the entropy agree with the relation (\ref{behae}).
These expressions hold even for the extremal black hole.
These results are consistent with (\ref{1stBTZ}).

The partition function of $AdS_3$ can be obtained in the similar way.
The semiclassical partition function of $AdS_3$ is given 
\bea
Z_{\rm semi.}^{\rm AdS}
= \exp\left(\frac{\beta_E}{8G_3}\right).
\label{parEADS}
\ena
By the continuation (\ref{conba}), the grand partition function 
becomes
\bea
Z^{\rm AdS} = \exp\left(\frac{1}{8G_3T}\right)\label{parADS}.
\ena
Because the Euclidian action proportional to the inverse temperature, 
the entropy vanishes.

\sect{AdS$_{\bf 3}$ Gravitational Instantons}\label{insta}
\noindent 
Euclidian $AdS_3$ is the upper half-space of $H^3$, and 
has an isometry $SL(2,{\bf C})$.
We introduce the Poincar\'e coordinates which is useful to observe the
identification by discrete subgroup of the isometry,
\bea
ds^2&=&\frac{\ell^2}{z^2}(dx^2 + dy^2 + dz^2)
= \frac{\ell^2}{z^2}(dw d\bar{w} +dz^2),\qquad z > 0,
\label{metP}
\ena
where $w \equiv x + y i$.
The action of $SL(2,{\bf C})$ can be expressed in terms of quaternion
\bea
q \equiv x  + y i + z j.
\ena
The $SL(2,{\bf C})$ action becomes
\bea
q \mapsto (aq + b)\,(cq + d)^{-1},
\qquad
{\bf H} =
\pmatrix{
a&b\cr
c&d} \in SL(2,{\bf C}).
\ena

The metric of the BTZ black hole (\ref{metEBTZ}) 
can be constructed as a quotient of $H^3$~\cite{cartei}.
We introduce left and right Euclidian temperature $T_{E\pm}$ as
\bea
\beta_E = \frac{1}{T_E} 
&=&\frac{1}{2}\left(\frac{1}{T_{E+}}+\frac{1}{T_{E-}}\right),
\nonumber\\
i\ell\frac{\Omega_{E+}}{T_E}
&=&\frac{1}{2}\left(\frac{1}{T_{E+}}-\frac{1}{T_{E-}}\right).
\ena
We introduce coordinates
\bea
w&=&\left(\frac{r^2 - r^2_+}{r^2 + |r^2_-|}\right)^{1/2}
\exp\left[2\pi T_{E+}(\ell\phi + i\tau)\right],\nonumber\\
z&=&\left(\frac{r^2_+ + |r^2_-|}{r^2 + |r^2_-|}\right)^{1/2}
\exp\left[\pi T_{E+} (\ell\phi + i\tau) + 
\pi T_{E_-} (\ell\phi -  i\tau)\right],
\label{cooEBTZ}
\ena
where $-\infty < \phi < \infty$ and $r > r_+$.
These coordinates are also available for the extremal black hole
except the massless black hole,
contrasting to the Lorentzian black holes~\cite{banteizenB}.
To make the coordinate $\phi$ into a true angular coordinate,
we must identify $\phi \sim \phi + 2\pi$.
This identification is given by the identification
\bea
w&\sim&\exp(4\pi^2\ell T_{E+}) w,\nonumber\\
z&\sim&\exp\left[2\pi^2\ell (T_{E+} + T_{E-})\right] z.
\label{ideBTZ}
\ena
The identifications implied by thermal ensemble are automatically 
taken into account as
\bea
\tau&\sim&\tau +\frac{1}{T_E},\nonumber\\
\phi&\sim&\phi - \frac{\Omega_{E+}}{T_E}.\label{ideBTZ2}
\ena
Note that the Euclidian BTZ black hole whose imaginary time identified 
has topology ${\bf R}^2 \times S^1$, i.e. a solid torus,
where $(r-r_+)$ and $\tau$ are the radial and angular coordinates in 
${\bf R}^2$ respectively. 
The direction of the imaginary time is a contractible loop of the
solid torus, and the core is a circle which represent the horizon.
It means that we have only anti-periodic fermion in the time direction.
The boundary at infinity, i.e. $z=0$ is a torus.
It is useful to introduce the modular parameter of the torus
\bea
\tau_M \equiv \frac{i}{2\pi\ell T_{E+}} =
-\frac{\Omega_{E+}}{2\pi T_E} + \frac{i}{2\pi\ell T_E}.
\label{modul}
\ena
We introduce coordinates of the torus as
\bea
\zeta \equiv \frac{i}{2\pi}\log w = 
T_{E+}(i\ell\phi -\tau),
\ena
and the identification (\ref{ideBTZ}) and (\ref{ideBTZ2}) are given 
by the modular parameter as
\bea
\zeta \sim \zeta - \frac{1}{\tau_M}, \qquad \zeta \sim \zeta -1,
\ena
respectively.

We construct the Euclidian $AdS_3$ with the identification (\ref{ideBTZ2}).
We introduce coordinates
\bea
w&=&\left(\frac{r^2}{r^2+\ell^2}\right)^{1/2}
\exp\left(\frac{\tau}{\ell}-i\phi\right),
\nonumber\\
z&=&\left(\frac{\ell^2}{r^2+\ell^2}\right)^{1/2}
\exp\left(\frac{\tau}{\ell}\right),\label{cooEA}
\ena
where we regard $\tau$, $r$ and $\phi$ as imaginary time, radial and
angular coordinates respectively.
Using the coordinates, the metric (\ref{metP}) becomes
\bea
ds^2 = \left(1+\frac{r^2}{\ell^2}\right)d\tau^{2}+
\left(1+\frac{r^2}{\ell^2}\right)^{-1}dr^{2}+r^{2}d\phi^{2},
\label{metEA}
\ena
The identification (\ref{ideBTZ2}) means
\bea
w&\sim&\exp\left(\frac{1}{\ell T_{E+}}\right)w,\nonumber\\
z&\sim&\exp\left(\frac{1}{\ell T_{E}}\right)z.
\label{ideEADS}
\ena
Note that the Euclidian $AdS_3$ whose imaginary time identified 
has topology ${\bf R}^2 \times S^1$,
where $r$ and $\phi$ are the radial and angular
coordinates in ${\bf R}^2$ respectively.
The direction of the imaginary time is a noncontructible loop of the
solid torus. It means that we have periodic or anti-periodic
fermion in the time direction.
The boundary at infinity, i.e. $z=0$ is a torus.
The coordinates of the torus becomes
\bea
\zeta = \frac{i}{2\pi}\left(\frac{\tau}{\ell} - i\phi \right), 
\ena
and the identification (\ref{ideEADS}) and $\phi \sim \phi +2\pi$
are given by the modular parameter (\ref{modul}) as
\bea
\zeta \sim \zeta +\tau_M, \qquad \zeta \sim \zeta +1,
\ena
respectively.

It is remarkable that the two boundary tori of the Euclidian black
hole and the Euclidian $AdS_3$  are related with the S-transformation
of the modular transformation:
\bea
{\rm S}: \tau_M \mapsto -\frac{1}{\tau_M}.
\ena
In other words, if we have a boundary torus of a Euclidian $AdS_3$
with modular parameter $\tau_M$, 
by applying the S-transformation, we get the
boundary torus of the Euclidian black hole at the same temperature.
However S-transformation is the diffeomorphism of the boundary torus,
it exchanges the contractible and the noncontructible loops in the bulk.
We may also introduce the T-transformation of the modular transformation:
\bea
{\rm T}: \tau_M \mapsto \tau_M + 1.
\ena
If we have a boundary torus of a Euclidian black hole whose
modular parameter is $\tau_M' \equiv -1/\tau_M$,
by applying the T-transformation, namely $\tau_M' \mapsto \tau_M' + 1$,
we get the same boundary torus of another Euclidian black hole. 
This boundary torus is obtained by applying Dehn twists around the
contractible loop of the boundary torus of the original Euclidian
black hole.
This transformation corresponds to
\bea
(r_+,|r_-|) \mapsto (r_+,|r_-|+\ell ).
\label{CTs}
\ena
This interesting symmetry of a boundary torus was found in
\cite{cartei}.

Let us consider thermodynamics in asymptotic $AdS_3$~\cite{malstrA,rez}.
The boundary of a system at an inverse temperature $\beta$ in
asymptotic $AdS_3$ has topology $T^2$ when continued to the Euclidian
metric.
Because this system involves angular momentum, 
we fix the boundary condition by not only $T_E$ but also $\Omega_{E+}$
as grand canonical ensemble.
One of the gravitational instantons which fit this boundary
condition is an Euclidian $AdS_3$ whose imaginary time is identified.
It has topology of a solid torus. 
Its semiclassical partition function (\ref{parEADS}) can be written as
\bea
Z_{\rm semi.}^{\rm AdS}[\tau_M]
&=&\exp\left[-\frac{i\pi k}{2}(\tau_M-{\bar{\tau_M}})\right].
\label{parEADS2}
\ena
The Euclidian BTZ black hole is another solution which fits 
the boundary condition. It also has topology of a solid torus.
Its semiclassical partition function (\ref{parEBTZ}) can be written as
\bea
Z_{\rm semi.}^{\rm BTZ}[\tau_M]
&=&\exp\left[\frac{i\pi k}{2}
\left(\frac{1}{\tau_M}-\frac{1}{\bar{\tau_M}}\right)\right].
\label{parEBTZ2}
\ena

Note that these two gravitational instantons are 
related by the S-transformation.
This fact is not surprising, because we discussed above,
when we have a boundary torus of an Euclidian $AdS_3$ with a
modular parameter $\tau_M$, this torus is a boundary torus of
Euclidian black hole whose modular parameter is $-1/\tau_M$.
In other words, we have at least two solid tori which fit the boundary 
torus.
Moreover, when we have a boundary torus of the Euclidian black hole, 
applying the S$^{-1}$TS-transformation, 
we get the same boundary torus of another Euclidian black hole,
but the bulk gravitational instanton (\ref{parEBTZ2}) is invariant.
This symmetry is the same as that of (\ref{CTs}).
The gravitational instanton (\ref{parEADS2}) is invariant under
T-transformation.

Inspired by these symmetry,
we suspect that for each modular parameter $\tau_M$
there are an $SL(2,{\bf Z})$ family of gravitational
instantons~\cite{malstrA} with partition functions
\bea
\exp\left[-\frac{i\pi k}{2}
\left(\frac{a\tau_M+b}{c\tau_M+d}-\frac{a\bar{\tau_M}+b}{c\bar{\tau_M}+d}
\right)\right],\qquad
\pmatrix{
a&b\cr
c&d} &\in& SL(2,{\bf Z}),
\ena
and these can be constructed beginning with the identifications
on $H^3$ by
\bea
{\bf H} = 
\pmatrix{
\exp\left(-i\pi\frac{a\tau_M+b}{c\tau_M+d}\right)&0\cr
0&\exp\left(i\pi\frac{a\tau_M+b}{c\tau_M+d}\right)}\in SL(2,{\bf C}).
\ena
Following this conjecture, the semiclassical partition function can be 
represented as
\bea
Z_{\rm semi.}[\tau_M] = \sum_{\rm family}
\exp\left[-\frac{i\pi k}{2}
\left(\frac{a\tau_M+b}{c\tau_M+d}-\frac{a\bar{\tau_M}+b}{c\bar{\tau_M}+d}
\right)\right],
\label{famil}
\ena
which has the modular invariance.
To derive the grand partition function, we must apply analytic
continuation of the angular momentum (\ref{conba}) to the
semiclassical partition function.
The grand partition function becomes
\bea
Z[T,\Omega_+] = \sum_{\rm family}
\exp\left[-\frac{i\pi\ell}{8G_3}
\left(\frac{ia+2\pi\ell T_+ b}{ic+2\pi\ell T_+ d}-
\frac{ia-2\pi\ell T_- b}{ic-2\pi\ell T_- d}
\right)\right],
\label{parfa}
\ena
where $T_{\pm}$ are given from Euclidian left and right temperature by 
the analytic continuation as
\bea
\frac{1}{T_{\pm}} = \frac{1}{T} \pm \frac{\ell\Omega_+}{T}.
\ena 
This expression includes the grand partition functions of $AdS_3$
(\ref{parADS}) and that of the black hole (\ref{parBTZ}). 
The expression (\ref{parfa}) includes unknown gravitational instantons
which give imaginary grand partition functions.
The grand partition function (\ref{parfa}) reaches that of $AdS_3$ 
in low temperature limit and that of the black hole in high temperature limit.
Because the BTZ black hole has positive specific heat,
it is stable at least locally.
In sufficiently  low temperature the black holes decays into $AdS_3$
and sufficiently high temperature $AdS_3$ decays into the black hole.
We can regard this phenomenon as phase transition between $AdS_3$
and the black hole. This phase transition becomes sharper 
when $k$ is large~\cite{malstrA}.

For the extremal black hole,
the partition function is also given by (\ref{parBTZ}),
and the behavior of (\ref{parBTZ}) may seem unclear because as $T$ reaches
zero, at the same time, $1-\Omega_+^2\ell^2$ reaches zero.
However, we see that the extremal black hole does not dominate.
This fact means that the Extremal black hole is unstable,
and decays into $AdS_3$~\cite{hawhuntey}.

\sect{Conformal Field Theory on the Boundary}\label{cft}  
\noindent
The class of boundary conditions which keep asymptotic $AdS_3$
generates conformal symmetry on the boundary~\cite{brohen}.
The generalization to $AdS_3$ supergravity~\cite{achtow} 
was discussed in~\cite{couhen}.
For $AdS_3$ we have four anti-periodic Killing spinors, two for 
each two-dimensional irreducible representation of the gamma-matrices.
We have the super-Virasoro algebra for each representation
with central charge $ c = 3\ell/2G_3$.
The super-Virasoro algebra implies bounds for the zero modes of
Virasoro generators.
We regard $AdS_3$ as the ground state which satisfies both bounds:
\bea
K^+_0|0_{NS}\rangle_L&=&0,\nonumber\\ 
K^-_0|0_{NS}\rangle_R&=&0,
\ena
where $K^{\pm}_0$ are the zero modes of the Virasoro generator.
Let us call this ground state as Neveu-Schwartz (NS) vacuum.

For the massless black hole, we have two periodic Killing spinors,
one for each two-dimensional irreducible representation of the gamma-matrices.
We have a set of the Ramond type super-Virasoro algebra.
The Ramond type super-Virasoro algebra implies bounds for the zero
modes of Virasoro generators.
We regard massless black hole as the ground state which satisfies
both bounds as
\bea
L^+_0|0_{R}\rangle_L&=&0,\nonumber\\ 
L^-_0|0_{R}\rangle_R&=&0.
\ena
Let us call this ground state as Ramond (R) vacuum.
For the extremal black hole, we have one periodic Killing spinor.
It means that the state which correspond to the extremal black hole
is killed by, say, the right-moving supercharge.

The above discussion of the ground states leads us to impose a 
relations between the eigenvalues of the generators and 
the mass and angular momentum of black holes:
\bea
H |{\rm state}\rangle&=&
\frac{1}{\ell}\left(L^+_0 + L^-_0\right)
|{\rm state}\rangle
= M |{\rm state}\rangle,\nonumber\\
P |{\rm state}\rangle&=&
\left(L^+_0 - L^-_0\right)|{\rm state}\rangle
= J |{\rm state}\rangle,\label{massR}
\ena
for R-sector and
\bea
H |{\rm state}\rangle&=&
\frac{1}{\ell}\left(K^+_0 + K^-_0-\frac{c}{12}\right)
|{\rm state}\rangle
= M |{\rm state}\rangle,\nonumber\\
P |{\rm state}\rangle&=&
\left(K^+_0 - K^-_0\right)|{\rm state}\rangle
= J |{\rm state}\rangle,\label{massNS}
\ena
for NS-sector.
With these relations, we find that R vacuum have $M = J = 0$ 
and states of the extremal black hole satisfies $M\ell = |J|$.
We also find that NS vacuum has $M = -(8G_3)^{-1}$ and $J = 0$. 

Let us recall that the BTZ black hole is emerged as the near horizon
geometry of the metric of the D-brane bound states.
The conformal field theory lives at infinity which emerges from 
the asymptotic symmetry and the D1-brane worldvolume theory
have the same central charge using the relation $c = 3\ell/2G_3 = 6k$.
When we consider these two conformal field theories are identical,
we expect that we can interpret
thermal phase transition of the bulk gravity in terms of the conformal 
field theory.
We have expectation that the partition function of the conformal field
theory is the same as that of bulk gravitational instantons~\cite{witH}.

Let us consider Euclidian conformal field theory lives on the 
complex plane with coordinates $w$.
The action of the global $SL(2,{\bf C})$ on $w$ is
\bea
w \mapsto \frac{aw+b}{cw+d},\qquad
\pmatrix{
a&b\cr
c&d\cr}\in SL(2,{\bf C}).
\ena
$H^3$ corresponds to the $SL(2,{\bf C})$ invariant vacuum.
We introduce coordinates of a torus as
\bea
\zeta = \frac{i}{2\pi} \log w = \tau + i \sigma,
\ena
and identify the coordinate as
\bea
\zeta \sim \zeta + \frac{1}{2\pi\ell T_E} +
i \frac{\Omega_{E+}}{2\pi T_E} = \zeta - i\tau_M.
\ena
This identification is consistent as finite temperature conformal
field theory.
Inspired by the $SL(2,{\bf Z})$ family of gravitational instantons,
we set the partition function on the torus to be modular invariant.
Unfortunately, the exact partition function of the conformal field theory 
on (\ref{targe}) is unknown, but we can expect that to observe the 
phase transition it is enough to use that of free theory.
When we have a bosonic sector only, the partition function becomes
modular invariant. However, we have a fermionic sector.
Because the modular transformation mix the boundary condition for
fermions, we must add spin structures on the torus to keep the
modular invariance~\cite{sup}.
In our discussion it is enough to take four sector of the spin structure:
$(+-;+-)$, $(++;++)$, $(--;--)$, $(-+;-+)$,
where $+$ and $-$ referring to periodic or anti-periodic respectively,
and the first pair of signs referring to left movers and the second
pair of signs to right movers,
and the first sign in each pair referring to the behavior 
in the $\sigma$ direction and the second referring to the
behavior in the $\tau$ direction.
The total partition function becomes modular invariant.
We denote the partition functions of each sector as
$Z_{E+-}$, $Z_{E++}$, $Z_{E--}$ and $Z_{E-+}$ respectively.

The partition functions are given as
\bea
Z_E&=&{\rm Tr} \exp\left(-\frac{1}{T_E}H+
i \frac{\Omega_{E+}}{T_E}P\right)
\ena
where
$H$ is the Hamiltonian and $P$ is the momentum which are defined in
(\ref{massR}) and (\ref{massNS}).
We define the functions for later convenience
\bea
f_1(q)&=&
q^{1/24}\prod^{\infty}_{n=1}\left(1-q^n\right),\nonumber\\
f_2(q)&=& 
\sqrt{2}q^{1/24}\prod^{\infty}_{n=1}\left(1+q^n\right),\nonumber\\
f_3(q)&=&
q^{-1/48}\prod^{\infty}_{n=1}\left(1+q^{n-1/2}\right),\nonumber\\
f_4(q)&=&
q^{-1/48}\prod^{\infty}_{n=1}\left(1-q^{n-1/2}\right),
\ena
where $q = \exp(2\pi i\tau_M)$.
They have the relations
\bea
f_1(q)&=&\left(\frac{i}{\tau_M}\right)^{1/2} f_1(\tilde{q}),\nonumber\\
f_2(q)&=&f_4(\tilde{q}),\nonumber\\
f_3(q)&=&f_3(\tilde{q}),
\ena
where $\tilde{q} = \exp(-2\pi i/\tau_M)$.
We define a function which is contained in all sector as
\bea
f_b(q)&=&\frac{1}{|f_1(q)|^2}({\rm Im} \tau_M)^{-1/2}
\ena
This function is the contribution from bosons and 
invariant under both of S-transformation and T-transformation.

The partition function of $(+-;+-)$ sector becomes
\bea
Z_{E+-}&=&{\rm Tr}\exp\left[-\frac{1}{\ell T_E}(L^+_0+L^-_0) 
+i \frac{\Omega_{E+}}{T_E}(L^+_0-L^-_0)\right]\nonumber\\
&=&\left(f_b(q)\right)^{2c/3}|f_2(q)|^{4c/3}.\label{par+-}
\ena
In this expression, we took account the degeneracy of the R vacuum.
This partition function is not invariant under S-transformation,
and invariant under T-transformation.
The partition function of $(--;--)$ sector becomes
\bea
Z_{E--}&=&{\rm Tr}\exp\left[-\frac{1}{\ell T_E}
(K^+_0 + K^-_0 -\frac{c}{12}) 
+i \frac{\Omega_{E+}}{T_E}(K^+_0 - K^-_0)\right]
\nonumber\\
&=&\left(f_b(q)\right)^{2c/3}|f_3(q)|^{4c/3}.\label{par--}
\ena
This partition function is invariant under S-transformation, and not
invariant under T-transformation.
In the finite temperature field theory,
the boundary condition for fermions are automatically imposed to be
anti-periodic.
To derive the partition function whose fermions are periodic,
we use the fermion number operator $(-)^{F}$.
The partition function of $(-+;-+)$ sector becomes
\bea
Z_{E-+}&=&{\rm Tr}(-)^F\exp\left[-\frac{1}{\ell T_E}
(K^+_0 + K^-_0  -\frac{c}{12}) 
+i \frac{\Omega_{E+}}{T_E}(K^+_0 - K^-_0)\right]
\nonumber\\
&=&\left(f_b(q)\right)^{2c/3}|f_4(q)|^{4c/3}.\label{par-+}
\ena
This partition function is not invariant under neither of 
S-transformation and T-transformation.
This partition function is transformed to (\ref{par+-}) under
S-transformation and to (\ref{par--}) under T-transformation.
The partition function of $(++;++)$ sector is
\bea
Z_{E++}&=&{\rm Tr}(-)^F\exp\left[-\frac{1}{\ell T_E}
(L^+_0+L^-_0) 
+i \frac{\Omega_{E+}}{T_E}(L^+_0-L^-_0)\right]
\nonumber\\
&=&0.\label{par++}
\ena
The reason why this partition function vanish is that the degenerate
R vacuum have fermion number and they cancel themselves,
as is well known in the GSO projection.

We have obtained four sectors of the partition function and 
they relate with each other by the modular transformation.
A construction of the partition function of 
the conformal field theory on the boundary of the BTZ black hole 
was discussed in terms of $SU(2)$ WZW model~\cite{banbroort}.
In this article, we have construct the partition function of the 
conformal field theory from another point of view.
The partition function of the Euclidian conformal field theory is the sum of
these partition functions
\bea
Z_{\rm ECFT} = Z_{E+-} + Z_{E--} + Z_{E-+}.
\label{ECFT}
\ena
In the following discussion, we observe our expectation that
the partition function is the same as that of bulk gravitational
instantons, namely
\bea
Z_{\rm ECFT} = Z_{\rm semi.}.
\ena

To obtain the grand partition functions,
we apply the continuation of the angular momentum
(\ref{conba}) to the partition functions of the Euclidian conformal
field theory.
This continuation is given by
\bea
q&\mapsto&q_L = \exp\left(-\frac{1}{\ell T_+}\right),\nonumber\\
\bar{q}&\mapsto&q_R = \exp\left(-\frac{1}{\ell T_-}\right).
\ena

Let us consider the nonextremal situation, namely $0 \ge \ell\Omega_+ > -1$.
To observe the behavior of partition functions in low temperature, 
we expand partition functions around the point
\bea
q_L = q_R = 0.
\ena
In low temperature limit,
the partition functions of each sector reach\foot{We discard
logarithmic contributions.}
\bea
Z_{+-}&\rightarrow&1,\\
Z_{--}&\rightarrow&(q_Lq_R)^{-c/24}
= \exp\left(\frac{1}{8G_3 T}\right),\\
Z_{-+}&\rightarrow&(q_Lq_R)^{-c/24}
= \exp\left(\frac{1}{8G_3 T}\right).
\ena
It is interesting that the partition functions of gravitational
instantons emerge in this limit.
We find that $Z_{+-}$ gives the partition function of the massless 
black hole and $Z_{--}$ and $Z_{-+}$ gives the partition function of $AdS_3$.
Moreover, a spin structure of the conformal field theory restrict the spin 
structure of bulk instantons~\cite{witH}.
As stated in section \ref{insta}, $AdS_3$ admit both spin structure in 
the time direction and anti-periodicity in the spatial direction.
This fact is consistent with the result that $Z_{--}$ and $Z_{-+}$ give
the partition function of $AdS_3$.
The massless black hole admits anti-periodicity in the time direction 
and periodicity in spatial direction. This fact is also consistent
with the result that $Z_{+-}$ gives the partition function of the
massless black hole.

To observe the behavior of partition functions in high temperature, 
we expand partition functions around the point
\bea
q_L = q_R = 1.
\ena
We introduce the parameter for convenience
\bea
\tilde{q}_L&=&\exp\left(-4\pi^2\ell T_+\right),\\
\tilde{q}_R&=&\exp\left(-4\pi^2\ell T_-\right).
\ena
In high temperature limit, 
the partition functions of each sector reach
\bea
Z_{+-}&\rightarrow&(\tilde{q}_L\tilde{q}_R)^{-c/24}
\approx \exp\left(\frac{\ell\pi^2}{4G_3(1-q_L)}+
\frac{\ell\pi^2}{4G_3(1-q_R)}\right),
\label{Z+-high}\\
Z_{--}&\rightarrow&(\tilde{q}_L\tilde{q}_R)^{-c/24}
\approx \exp\left(\frac{\ell\pi^2}{4G_3(1-q_L)}+
\frac{\ell\pi^2}{4G_3(1-q_R)}\right),
\label{Z--high}\\
Z_{-+}&\rightarrow&1,
\ena
where we approximate as $\log q \approx q-1$ for $q \rightarrow 1$. 
In high temperature limit, we can expand $q_L$ and $q_R$ by 
$T_{\pm}$, and we may approximate above expressions further 
\bea
q_L&\approx&1-\frac{1}{\ell T_+},\nonumber\\
q_R&\approx&1-\frac{1}{\ell T_-}.\label{qLqRa}
\ena
The partition functions $Z_{+-}$ and $Z_{--}$ reaches
\bea
Z_{+-}&\approx&\exp\left(\frac{\pi^2\ell^2}{2G_3}
\frac{T}{1-\Omega^2_+\ell^2}\right),\\
Z_{--}&\approx&\exp\left(\frac{\pi^2\ell^2}{2G_3}
\frac{T}{1-\Omega^2_+\ell^2}\right).
\ena
We find that $Z_{-+}$ gives the partition function of the
massless black holes, and $Z_{+-}$ and $Z_{--}$ gives the 
partition function of the black hole.
It is interesting that both of $Z_{+-}$ and $Z_{--}$ give the
partition function of the black hole.
It means that black holes do not have supersymmetry, and admit both  
spin structure in the spatial direction.
Moreover, these two sectors are related by the S$^{-1}$TS-transformation.

We can obtain the microscopic entropy from the partition functions
(\ref{Z+-high}) and (\ref{Z--high}) as degeneracy of states.
We may rewrite $Z_{+-}$ as
\bea
Z_{+-}&=&{\rm Tr}q_L^{L^+_0} q_R^{L^-_0}
= \sum^{\infty}_{n_L=0}\sum^{\infty}_{n_R=0}d_{n_L,n_R}q_L^{n_L}q_R^{n_R}
\nonumber\\
&=&\sum_{m}\sum_{j}d_{m,j}
\exp\left[- \beta m + \beta \Omega_+ j \right].
\ena
where $n_L$ and $n_R$ are level of states which
are written by $m$ and $j$ using (\ref{massR}).
The degeneracy of states is given as
\bea
d_{N_L,N_R}&=&\frac{1}{(2\pi i)^2}
\oint dq_Ldq_R q_L^{-N_L-1}q_R^{-N_R-1}Z_{+-}.
\ena
We obtain the entropy as
\bea
S = \log d_{N_L,N_R} = \frac{A}{4G_3}.
\ena
This reproduce the entropy (\ref{behae}) correctly.
This saddle point evaluation is available when
$N_L, N_R \gg c$, namely $G_3 M \gg 1$.
The same argument holds for $Z_{--}$.

In low temperature limit $Z_{--}$ dominates and it 
reproduces partition function of $AdS_3$ and
in high temperature limit $Z_{+-}$ or $Z_{--}$ dominate, 
and they reproduce the partition function of the black hole.
These phenomena reproduce the phase transition of 
the bulk gravity as expected.

Let us consider the extremal limit where $\ell\Omega_+ = -1$ 
and $T_- = 0$.
In this limit, the parameter $q_R$ reaches zero.
On the other hand, the parameter $q_L$ reaches
\bea
q_L \rightarrow \exp\left(-\frac{\pi}{\sqrt{2G_3 M}}\right).
\ena
Thus, when $G_3 M \gg1$ we have another limit than low temperature
limit.
We have interest in this limit. To observe the behavior of
partition functions in this limit,
we expand partition functions around the point
\bea
q_L = 1, \qquad q_R = 0.
\ena
In this limit,
the partition function of each sector reaches
\bea
Z_{+-}&\rightarrow&\tilde{q}^{-c/24}_L
\approx \exp\left(\frac{\ell\pi^2}{4G_3(1-q_L)}\right),\\
Z_{--}&\rightarrow&q^{-c/24}_R\tilde{q}^{-c/24}_L
\approx
q^{-\ell/16G_3}_R\exp\left(\frac{\ell\pi^2}{4G_3(1-q_L)}\right),
\label{Z--ext}\\
Z_{-+}&\rightarrow&q^{-c/24}_R \approx q^{-\ell/16G_3}_R,
\ena
where we approximate as $\log q_L \approx q_L-1$ for $q_L \rightarrow 1$.
Using the approximation (\ref{qLqRa}), we obtain
\bea
Z_{+-}&\approx&
\exp\left(\frac{\ell^2\pi^2T_+}{4G_3}\right),\\
Z_{--}&\approx&
\exp\left(\frac{1}{8G_3 T}\right)
\exp\left(\frac{\ell^2\pi^2T_+}{4G_3}\right),\\
Z_{-+}&\approx&
\exp\left(\frac{1}{8G_3 T}\right).
\ena
We find that $Z_{+-}$ gives the partition function of the
extremal black holes and $Z_{-+}$ gives the partition function of
$AdS_3$.
$Z_{--}$ also gives the partition function of $AdS_3$ 
in this limit, namely the first factor dominates. 
In this limit $Z_{--}$ dominates
and it reproduces the partition function of $AdS_3$.
This fact reproduces that the extremal BTZ black hole is unstable,
and decays into $AdS_3$.
However, there seems to be an inconsistency that 
the microscopic entropy of $Z_{--}$ is the same as that of 
the extremal black hole because of the behavior (\ref{Z--ext}),
nevertheless $AdS_3$ has no entropy in
Euclidian gravity approach.
This inconsistency may needs further investigations.

\sect{Discussions}
We have discussed $AdS_3$ gravity in Euclidian quantum gravity
approach.
We constructed the semiclassical partition function which includes
all members of the $SL(2,{\bf Z})$ family of gravitational instantons.
In high temperature limit the black hole dominates and in
low temperature limit $AdS_3$ dominates. 
This phenomenon was regarded as thermal phase transition.

We constructed partition function of 
a conformal field theory on the boundary of $AdS_3$ gravity.
Inspired by the $SL(2,{\bf Z})$ family, we constructed the partition
function such that $SL(2,{\bf Z})$ invariant in accordance with the
spin structure.
The construction of the partition function of the conformal field
theory on the boundary of black holes have been considered in terms 
of $SU(2)$ WZW model in~\cite{banbroort}.
Our interest was that whether the partition function of the conformal 
field theory on the boundary reproduce not only the behavior of 
black holes, but also the thermal phase transition in bulk gravity.
We have discussed the thermal behavior of the
conformal field theory which is obtained 
as low energy effective theory of the D1-brane worldvolume theory
using the partition function of free theory.
More rigid discussion of the thermal behavior needs to understand
the partition function on the deformation of the symmetric product.

It was shown that the partition function of the
nonextremal black hole and that of $AdS_3$ emerge as high 
temperature and low temperature limit of the partition function 
of the conformal field theory respectively.
These results reproduce the phase transition in the bulk gravity.
In the extremal limit, the partition function of the conformal 
field theory becomes the form of the partition function of $AdS_3$.
It is consistent with the fact that the extremal BTZ black hole is
unstable thermodynamically.
However, the microscopic entropy from the partition function of
the conformal field theory seems to exist.
This phenomena may need further investigation.

Let us recall the correspondence between $AdS_5$ gravity and the
conformal field theory on the boundary which was discussed
in~\cite{witH}.
We have two gravitational instantons; the one is $AdS_5$, 
and the other is anti-de Sitter-Schwarzschild black hole.
The anti-de Sitter-Schwarzschild black hole does not admit
periodicity of fermions in the time direction, the phase transition
occurred in the partition function whose periodicity of fermions are
anti-periodic in time direction,
namely 
\bea
Z^{\rm AdS}_{\rm semi.} + Z^{\rm BH}_{\rm semi.} = {\rm Tr}e^{-\beta H}
\ena
In the correspondence between $AdS_3$ gravity and the conformal field
theory which was discussed in this article, 
we have four sector of the partition function of the conformal field
theory because of the spin structure.
As shown above, it seems that the partition function $Z_{--}$
reproduces $AdS_3$ in low temperature and the black hole in high
temperature, however $Z_{--}$ itself is not modular invariant.
Thus, in this article, we consider the correspondence of 
modular invariant partition functions
\bea
Z_{\rm semi.} = Z_{\rm ECFT}
\ena
where the left hand side is the sum of the family of gravitational
instantons and the right hand side is the sum of the three sectors with
different spin structure.
To investigate the intermediate region of temperature,
which is nontrivial in terms of both of gravitational instantons and
conformal field theory, may make clear the correspondence.

\vspace{0.7cm}
\noindent
{\Large\bf Acknowledgments}\\
\\
\noindent
I am grateful to K. Furuuchi and T. Nakatsu for fruitful
discussions. 
I would like to thank J-G. Zhou for helpful 
suggestions and reading the manuscript.
I would also like to thank N. Ohta, M. Torii, H. Umetsu, 
N. Yokoi and Y. Yoshida for helpful suggestions.
\newpage


\end{document}